\documentclass[aps,prl,twocolumn]{revtex4}  
\usepackage{graphicx}  
\usepackage{dcolumn}   
\usepackage{bm}        
\usepackage{amssymb}   

\begin{document}

\widetext


\title{Recent Belle results related to $\pi-K$ interactions}
\author{Bilas Pal\\ (On behalf of the Belle Collaboration)\\ Brookhaven National Laboratory, Upton, NY \& University of Cincinnati, Cincinnati, OH}
 

\begin{abstract}
We report the recent results related to  $\pi-K$ interactions based on the data collected by the Belle experiment at the KEKB collider. This includes the branching fraction and $CP$ asymmetry measurements of $B^+\to K^+K^-\pi^+$ decay, search for the $\Lambda_c^+\to\phi p\pi^0$, $\Lambda_c^+\to P_s^+\pi^0$ decays, branching fraction measurement of $\Lambda_c^+\to K^-\pi^+p\pi^0$, first observation of  doubly Cabibbo-suppressed decay $\Lambda_c^+\to K^+\pi^-p$, and the measurement of CKM angle $\phi_3$ ($\gamma$) with a model-independent Dalitz plot analysis of $B^{\pm}\to DK^{\pm},D\to K_S^0\pi^+\pi^-$ decay.
\end{abstract}

\maketitle

\section{Introduction}
In this report, we present some recent results related to $\pi-K$ interactions based on the  data, collected by the Belle experiment at the KEKB $e^+e^-$ asymmetric-energy collider~\cite{KEKB}. 
(Throughout this paper charge-conjugate modes are implied.) 
The experiment
took data at center-of-mass energies corresponding
to several $\Upsilon(nS)$ resonances; the total data sample
recorded exceeds $1~{\rm ab}^{-1}$.

The Belle detector is a large-solid-angle magnetic
spectrometer that consists of a silicon vertex detector
(SVD), a 50-layer central drift chamber (CDC),
an array of aerogel threshold Cherenkov counters
(ACC), a barrel-like arrangement of time-of-flight
scintillation counters (TOF), and an electromagnetic
calorimeter comprised of CsI(Tl) crystals
(ECL) located inside a super-conducting solenoid
coil that provides a 1.5 T magnetic field. An iron
flux-return located outside of the coil is instrumented
to detect $K^0_L$
mesons and to identify muons
(KLM). The detector is described in detail elsewhere~\cite{Belle, svd2}.

\section{$CP$ asymmetry in $B^+\to K^+K^-\pi^+$ decays}
In the recent years, an unidentified structure has been observed by BaBar~\cite{Aubert:2007xb} and LHCb experiments~\cite{Aaij:2013bla, Aaij:2014iva} in the low $K^+K^-$
invariant mass spectrum of the $B^+\to K^+K^-\pi^+$ decays. The LHCb reported a nonzero inclusive $CP$ asymmetry of $-0.123\pm0.017\pm0.012\pm0.007$ and a large unquantified local $CP$ asymmetry in the same mass region. These results suggest that final-state interactions may contribute to $CP$ violation~\cite{Bhattacharya:2013boa, Bediaga:2013ela}. In this analysis, we attempt to quantify the $CP$ asymmetry and branching fraction as a function of the $K^+K^-$ invariant mass, using $711~{\rm fb^{-1}}$ of data, collected at $\Upsilon(4S)$ resonance~\cite{Hsu:2017kir}.

The signal yield is extracted by performing a two-dimensional unbinned maximum likelihood fit to the variables: the beam-energy constrained mass $M_{\rm bc}$  and the energy difference $\Delta E$. The resulting branching fraction and $CP$ asymmetry are
\begin{eqnarray*}
\mathcal{B}(B^+\to K^+K^-\pi^+)&=&(5.38\pm0.40\pm0.35)\times10^{-6},\\
A_{CP}&=&-0.170\pm0.073\pm0.017,
\end{eqnarray*}
where the quoted uncertainties are statistical and systematic, respectively.

\begin{figure}[htb]
\centering
\includegraphics[width=0.5\textwidth]{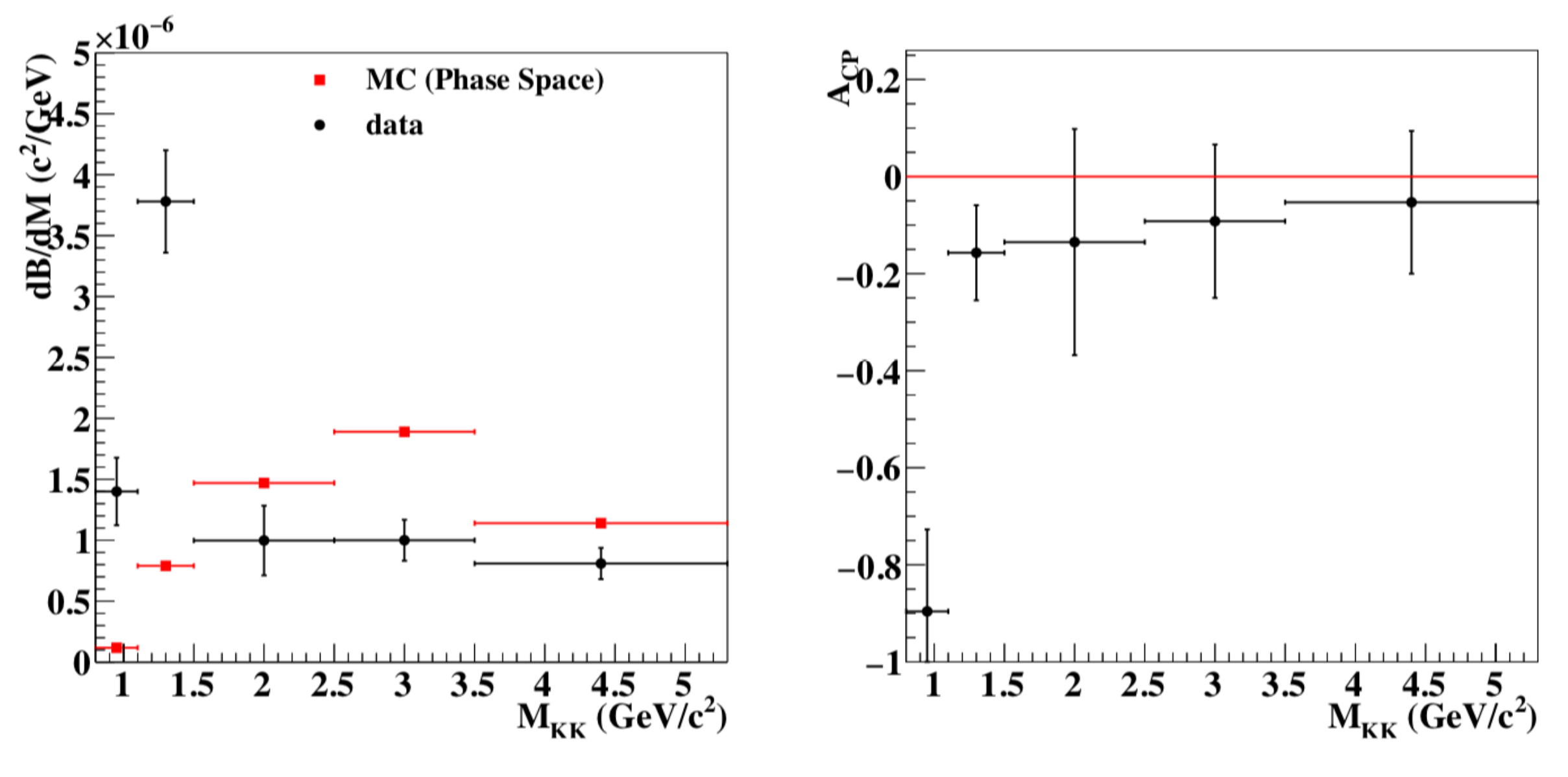}
\vskip -0.35cm
\caption{\small  Differential branching fractions~(left) and measured $A_{CP}$
~(right) as a function of $M_{K^+K^-}$. Each point is obtained from a two-dimensional fit with systematic uncertainty included. Red squares with error bars in the left figure show the expected signal distribution in a three-body phase space MC. Note that the phase space hypothesis is rescaled to the total observed $K^+K^-\pi^+$ signal yield.}
\label{fig:cp_all}
\end{figure}

To investigate the localized $CP$ asymmetry in the low $K^+K^-$
invariant mass region, we perform the 2D fit (described above) to extract the signal yield and $A_{CP}$ in bins of $M_{K^+K^-}$. The fitted results are shown in Fig.~\ref{fig:cp_all} and Table~\ref{tab:cp_all}. We confirm the excess and local $A_{CP}$ in the low  $M_{K^+K^-}$ 
 region, as reported by the LHCb, and quantify the differential branching fraction in each $K^+K^-$ invariant mass bin. We find a 4.8$\sigma$ evidence for a negative $CP$ asymmetry in the region $M_{K^+K^-}<1.1$ GeV/$c^2$. To understand the origin of the low-mass dynamics, a full Dalitz analysis from experiments with a sizeable data set, such as LHCb and Belle II, will be needed in the future.
\begin{table}[h!t!p!]
\centering
\caption{\small Differential branching fraction, and $A_{CP}$ for individual $M_{K^+K^-}$ bins. The first uncertainties are statistical and the second systematic.}
\vspace{0.2cm}
\begin{tabular}{c|cc}
\hline\hline
        $M_{K^+K^-}$ & $d\mathcal{B}/dM (\times 10^{-7})$ & $A_{CP}$\\
\hline
0.8--1.1 & $14.0\pm2.7\pm0.8$ & $-0.90\pm0.17\pm0.04$\\
1.1--1.5 & $37.8\pm3.8\pm1.9$ & $-0.16\pm0.10\pm0.01$\\
1.5--2.5 & $10.0\pm2.3\pm1.7$ & $-0.15\pm0.23\pm0.03$\\
2.5--3.5 & $10.0\pm1.6\pm0.6$ & $-0.09\pm0.16\pm0.01$\\
3.5--5.3 & $8.1\pm1.2\pm0.5$   & $-0.05\pm0.15\pm0.01$\\
\hline    \hline    
\end{tabular}
\label{tab:cp_all}
\end{table}

\section{Search for \mbox{\boldmath$\Lambda_c^+\to\phi p \pi^0$} and branching fraction measurement of  \mbox{\boldmath$\Lambda_c^+\to K^-\pi^+ p \pi^0$} }

\begin{figure}[htb]
\centering
\includegraphics[width=0.20\textwidth]{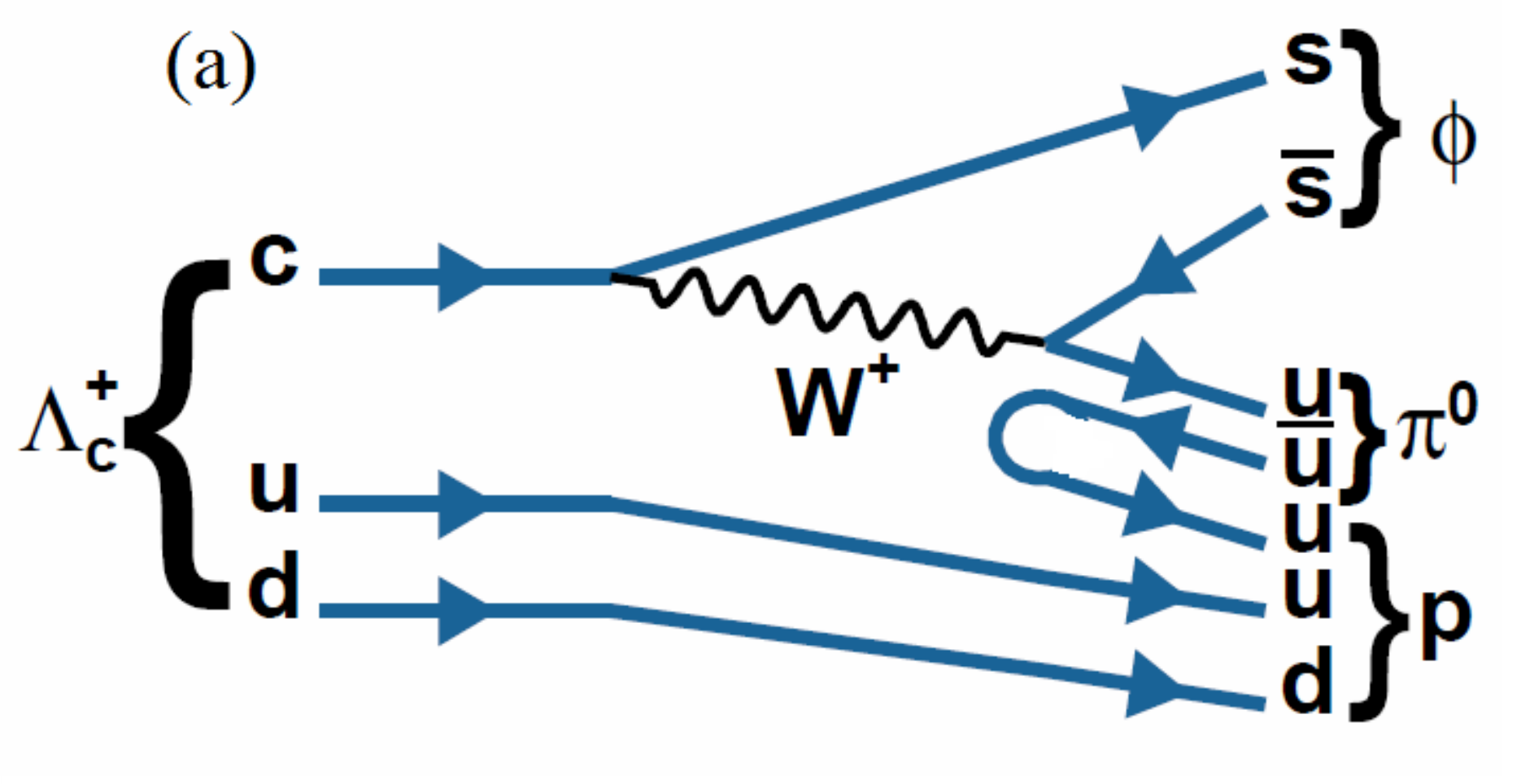}%
\includegraphics[width=0.20\textwidth]{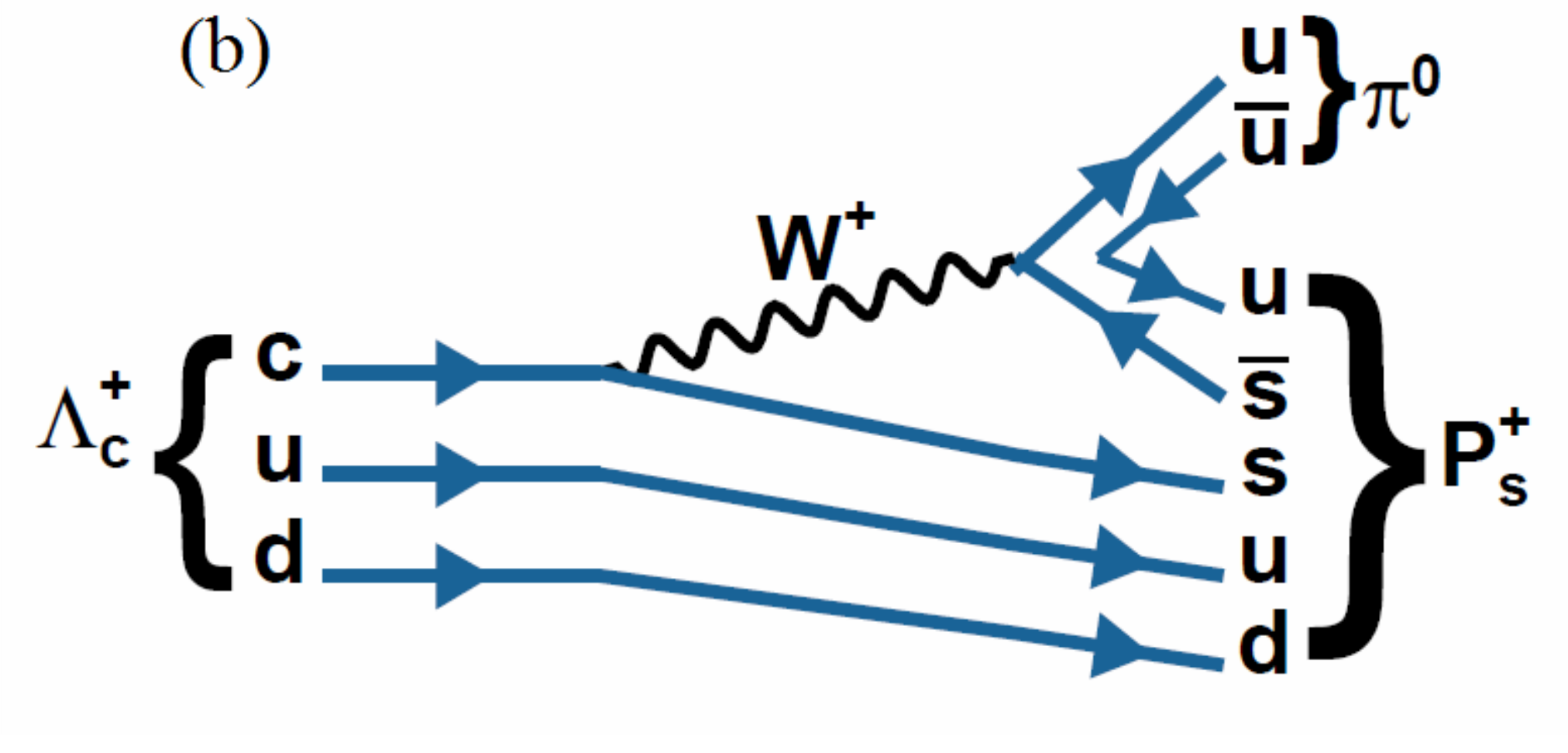}
\vskip -0.3cm
\caption{\small Feynman diagram for the decay (a) $\Lambda_c^+\to\phi p\pi^0$ and (b) $\Lambda_c^+\to P_s^+\pi^0$.}
\label{fig:Feynman}
\end{figure}

The story of exotic hadron  spectroscopy   begins  with the  discovery  of  the
$X(3872)$ by the Belle collaboration in 2003~\cite{Choi:2003ue}. Since then, many exotic $X\!Y\!Z$ states have been reported by Belle and other experiments~\cite{Agashe:2014kda}. Recent observations of two hidden-charm pentaquark  states $P_c^+(4380)$ and $P_c^+(4450)$ by the LHCb collaboration in the $J/\psi p$ invariant mass spectrum of the  $\Lambda_b^0\to J/\psi pK^- $ process~\cite{Aaij:2015tga} raises the question of whether a hidden-strangeness pentaquark  $P_s^+$, where the $c\bar{c}$ pair  in  $P_c^+$  is replaced by  an $s\bar{s}$ pair, exists~\cite{Kopeliovich:2015vqa, Zhu:2015bba, Lebed:2015dca}. The strange-flavor analogue of the $P_c^+$ discovery channel is the decay $\Lambda_c^+\to\phi p\pi^0$~\cite{Kopeliovich:2015vqa, Lebed:2015dca}, shown in Fig.~\ref{fig:Feynman} (a). The detection of a hidden-strangeness pentaquark could be possible through  the $\phi p$ invariant mass spectrum within this channel [see Fig.~\ref{fig:Feynman} (b)]
if the underlying mechanism creating the $P_c^+$ states also holds for $P_s^+$, independent of the flavor~\cite{Lebed:2015dca}, and only if  the mass of $P_s^+$ is less than $M_{\Lambda_c^+}-M_{\pi^0}$.  
In an analogous $s\bar{s}$ process of $\phi$ photoproduction $(\gamma p\to\phi p)$,  a forward-angle  bump structure at $\sqrt{s}\approx2.0$ GeV 
has been observed by  the LEPS~\cite{Mibe:2005er} and CLAS collaborations~\cite{Dey:2014tfa}.
However, this structure appears only at the most forward angles, 
which is  not   expected for  the decay of a resonance~\cite{Lebed:2015fpa}.

Previously, the decay $\Lambda_c^+\to\phi p\pi^0$ has not been studied by any experiment. Here, we report a search for this decay, using 915 $\rm fb^{-1}$ of data~\cite{Pal:2017ypp}. In addition, we search for the nonresonant decay $\Lambda_c^+\to K^+K^-p\pi^0$ and measure the branching fraction of the  Cabibbo-favored decay $\Lambda_c^+\to K^-\pi^+p\pi^0$.

\begin{figure}[h!tp]
\begin{center}
    \includegraphics[width=0.24\textwidth]{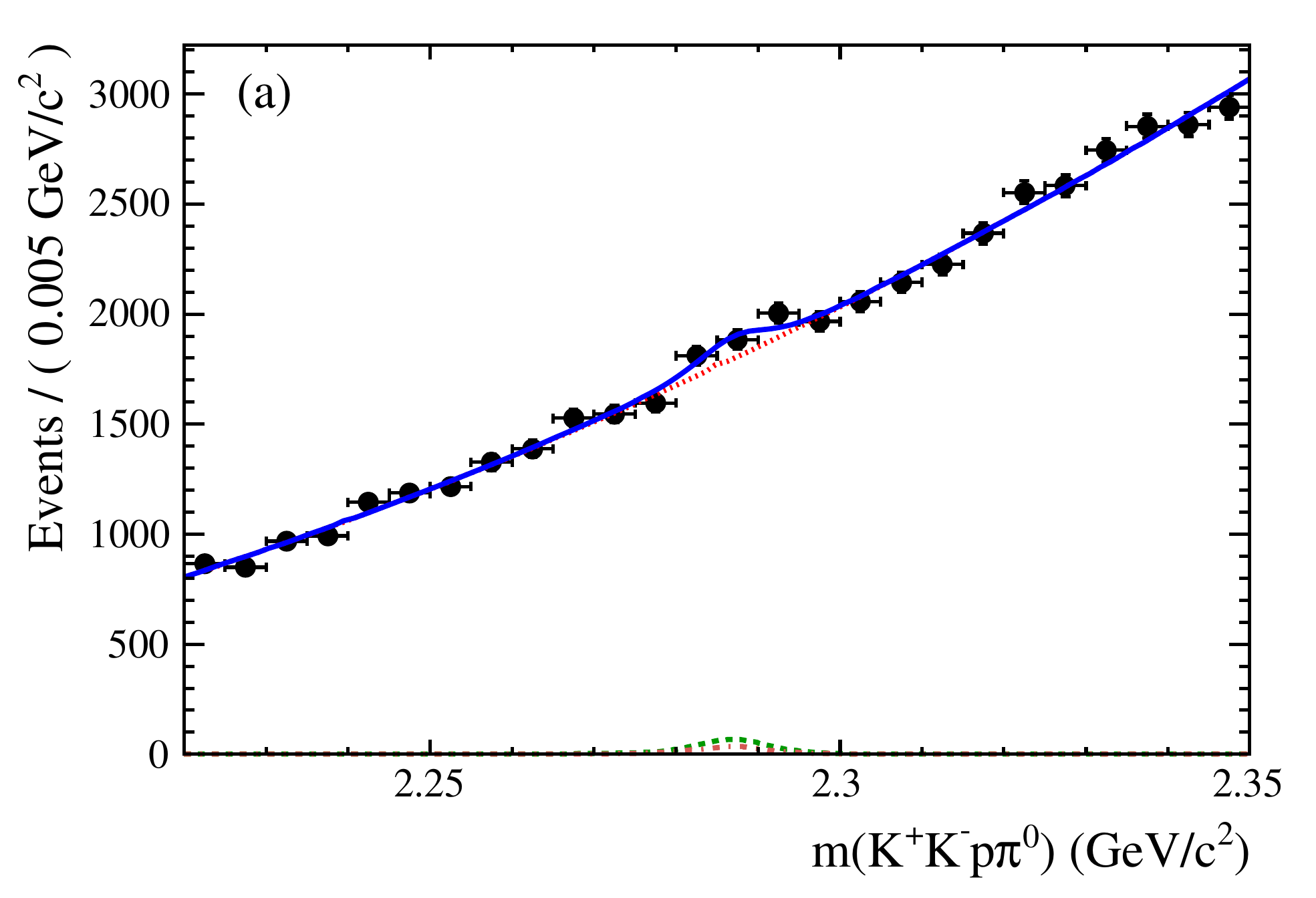}%
       \includegraphics[width=0.24\textwidth]{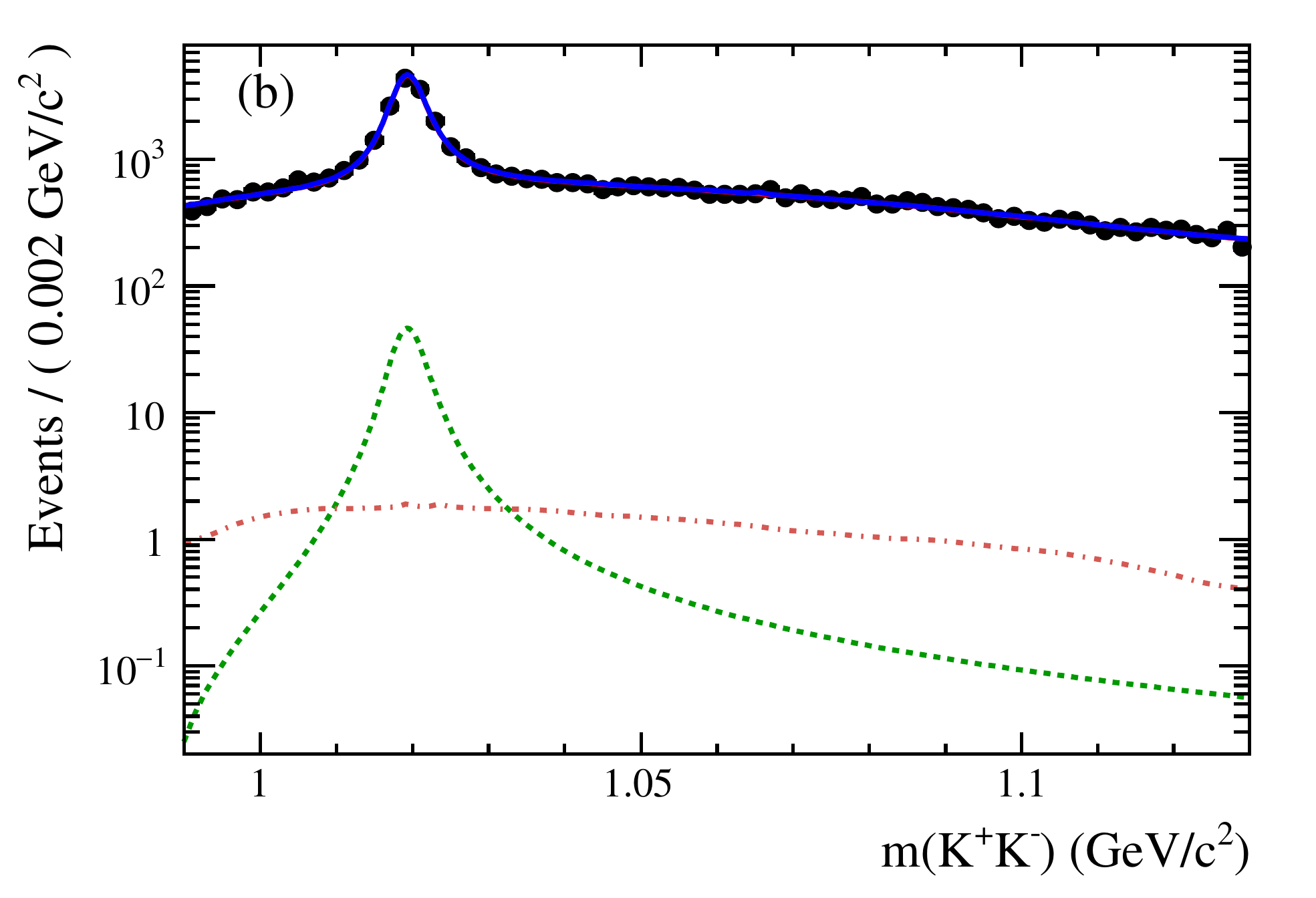}
\end{center}
\vskip -0.75cm
\caption{\small Projections of the 2D fit: (a) $m(K^+K^- p\pi^0)$ and (b) $m(K^+K^-)$. The points with the error bars are the  data, and the (red) dotted, (green) dashed and (brown) dot-dashed curves represent the combinatorial,  signal and nonresonant candidates, respectively, and (blue) solid curves represent the total PDF. The solid curve in (b) completely overlaps the curve for the combinatorial background.}
\label{fig:2dfit}
\end{figure}

In order to extract the signal yield, we perform a two-dimensional (2D) unbinned extended maximum likelihood fit to the variables $m (K^+K^-p\pi^0)$ and  $m(K^+K^-)$.  Projections of the fit result are shown in Fig.~\ref{fig:2dfit}. 
From the fit, we extract $148.4\pm61.8$ signal events, $75.9\pm84.8$ nonresonant events, 
and $7158.4\pm36.4$ combinatorial background events. 
The statistical significances are found to be 2.4 and 1.0 standard deviations for $\Lambda_c^+\to\phi p \pi^0$ and nonresonant $\Lambda_c^+\to K^+K^- p \pi^0$ decays, respectively. We use the well-established decay $\Lambda_c^+\to p K^-\pi^+$~\cite{Agashe:2014kda} as the normalization channel for the branching fraction measurements. 

Since the significances are below 3.0 standard deviations both for $\phi p\pi^0$ signal and $K^+K^-p\pi^0$ nonresonant decays, we set upper limits on their branching fractions  at 90\% confidence level (CL) using a Bayesian approach. 
The results are
\begin{eqnarray*}
\mathcal{B}(\Lambda_c^+\to \phi p\pi^0) &<& 15.3\times10^{-5} ,\\
\mathcal{B}(\Lambda_c^+\to K^+K^-p\pi^0)_{\rm NR} &<&6.3\times10^{-5} ,
\end{eqnarray*}
which are the first limits on these branching fractions. 

\begin{figure}[htb]
\centering
\includegraphics[width=0.425\textwidth]{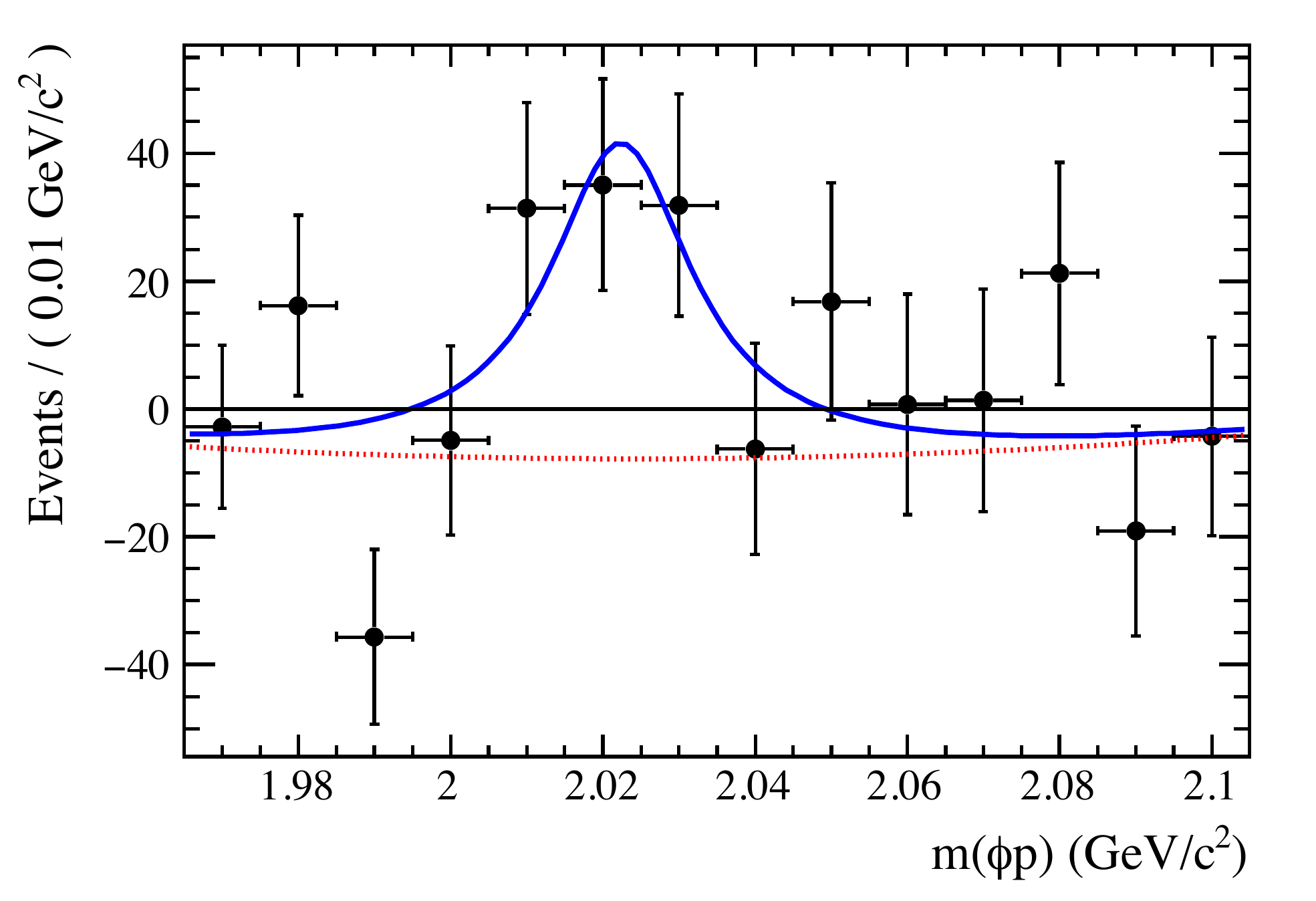}
\vskip -0.3cm
\caption{\small   The background-subtracted distribution of $m(\phi p)$ in the $\phi p\pi^0$ final state. The points with error bars are data, and   the (blue) solid line shows the total PDF. The (red) dotted curve shows the fitted phase space component (which has fluctuated negative).}
\label{fig:bkg-sub_dis}
\end{figure}

To search for a putative $P_s^+\to\phi p$ decay, we select $\Lambda_c^+\to K^+K^-p\pi^0$ candidates in which $m(K^+K^-)$ is within 0.020~GeV/$c^2$ of the   $\phi$ meson mass~\cite{Agashe:2014kda}
and plot the  background-subtracted $m(\phi p)$ distribution (Fig.~\ref{fig:bkg-sub_dis}). This distribution is obtained by performing 2D fits as discussed above in bins of $m(\phi p)$. 
The data shows no clear evidence for a $P_s^+$ state. 
We set an upper limit on the product branching fraction 
$\mathcal{B}(\Lambda_c^+\to P_s^+\pi^0) \times \mathcal{B}(P_s^+\to \phi p)$ by fitting the  distribution of Fig.~\ref{fig:bkg-sub_dis}
to the sum of a RBW function and a phase space 
distribution determined from a sample of simulated $\Lambda^+_c\to\phi p\pi^0$ 
decays. We obtain $77.6\pm28.1$ $P_s^+$ events from the fit, which gives an upper limit of 
\begin{eqnarray*}
\mathcal{B}(\Lambda_c^+\to P_s^+\pi^0) \times 
\mathcal{B}(P_s^+\to \phi p)  <  8.3\times 10^{-5}
\end{eqnarray*}
at 90\% CL. 
From the fit, we also obtain, 
$M_{P_s^+}=(2.025\pm 0.005)$~GeV/$c^2$ and 
$\Gamma_{P_s^+}=(0.022\pm 0.012)$~GeV,
where the uncertainties are statistical only.

\begin{figure}[h!tb]
\centering
\includegraphics[width=0.425\textwidth]{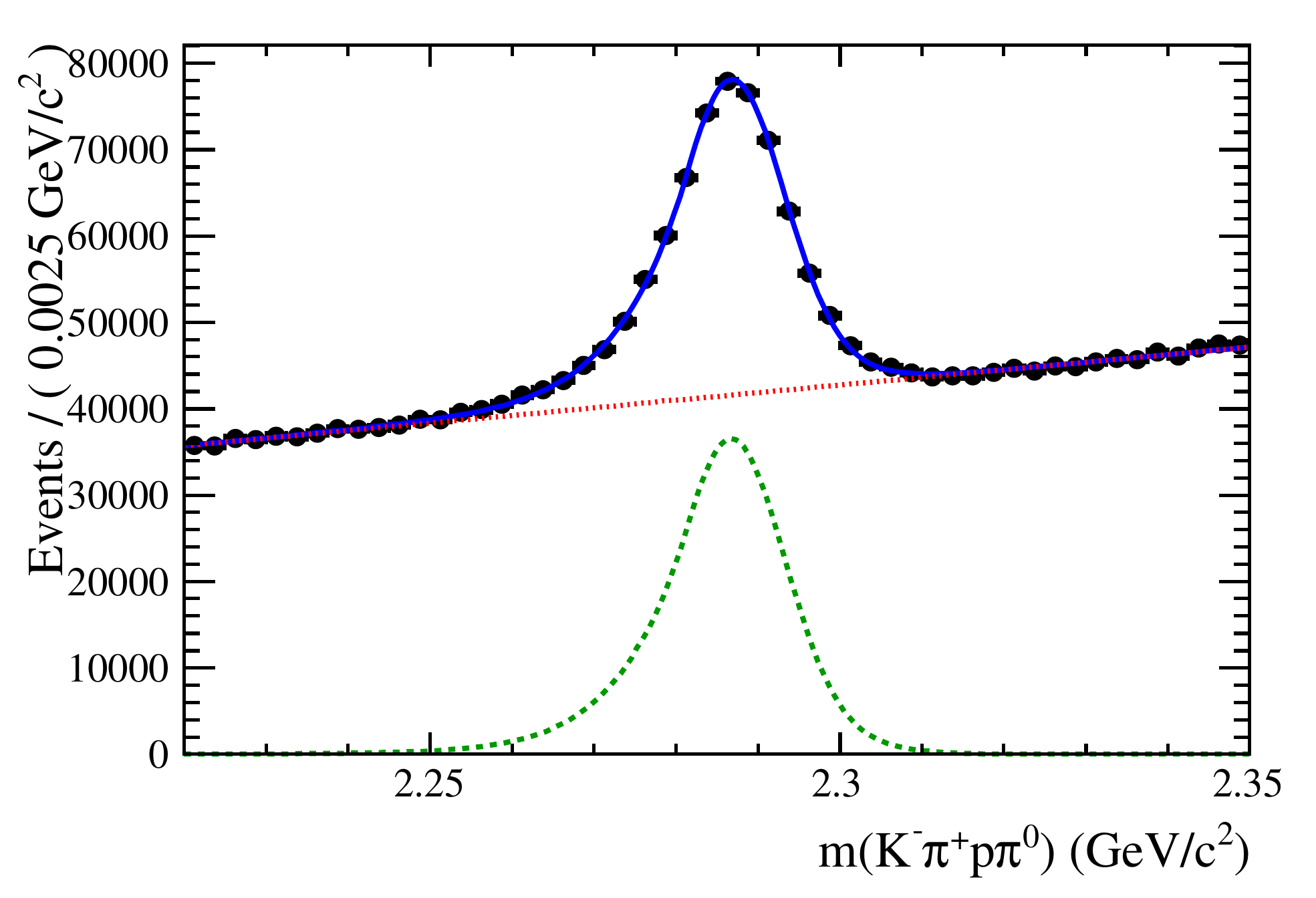}
\vskip -0.3cm
\caption{\small  Fit to the invariant mass distribution of $m(K^-\pi^+p\pi^0)$. The points with the error bars are the  data,  the (red) dotted and (green) dashed curves  represent the combinatorial and  signal  candidates, respectively, and (blue) curve represents the total PDF.  
}
\label{fig:invmass_control2_data}
\end{figure}

The high statistics decay  $\Lambda_c^+\to K^-\pi^+p\pi^0$ is used to adjust the data-MC differences in the $\phi p\pi^0$ signal and $K^+K^-p\pi^0$ nonresonant decays. For the $\Lambda_c^+\to K^-\pi^+p\pi^0$ sample, 
the mass distribution is plotted in Fig.~\ref{fig:invmass_control2_data}. We fit this distribution to obtain the signal yield. We find $242\,039\pm \,2342$ signal candidates and $472\,729\pm\,467$ background candidates. We measure the ratio of branching fractions,
\begin{eqnarray*}
\frac{\mathcal{B}(\Lambda_c^+\to K^-\pi^+p\pi^0)}{\mathcal{B}(\Lambda_c^+\to K^-\pi^+p)}=(0.685\pm0.007\pm 0.018),
\end{eqnarray*}
where the first uncertainty is statistical and the second is systematic. Multiplying this ratio by the world average value of $\mathcal{B}(\Lambda_c^+\to K^-\pi^+p)=(6.46\pm0.24)\%$~\cite{Amhis:2016xyh}, we obtain
\begin{eqnarray*}
\mathcal{B}(\Lambda_c^+\to K^-\pi^+p\pi^0)=(4.42\pm0.05\pm 0.12\pm0.16)\%,
\end{eqnarray*}
where the first uncertainty is statistical, the second is systematic, and the third reflects the uncertainty due to the branching fraction of the normalization decay mode. This is the most precise measurement of $\mathcal{B}(\Lambda_c^+\to K^-\pi^+p\pi^0)$ to date and is consistent with the recently measured value $\mathcal{B}(\Lambda_c^+\to K^-\pi^+p\pi^0)=(4.53\pm0.23\pm0.30)\%$ by the BESIII collaboration~\cite{Ablikim:2015flg}.

\section{Observation of the doubly Cabibbo-suppressed \mbox{\boldmath$\Lambda_c^+$} decay }
Several doubly Cabibbo-suppressed (DCS) decays of charmed mesons have been observed~\cite{Agashe:2014kda}. Their measured branching ratios with respect to the corresponding Cabibbo-favored (CF) decays play an important role in
constraining models of the decay of charmed hadrons and in the study of flavor- $SU(3)$ symmetry~\cite{Lipkin:2002za, Gao:2006nb}. On the other hand, because of the smaller production cross-sections for charmed baryons, DCS decays of charmed
baryons have not yet been observed, and only an upper limit, $\frac{\mathcal{B}(\Lambda_c^+\to pK^+\pi^-)}{\mathcal{B}(\Lambda_c^+\to pK^-\pi^+)}<0.46\%$ at 90\% CL, has been reported by the FOCUS Collaboration~\cite{Link:2005ym}. Here we present the first observation of the DCS decay $\Lambda_c^+\to pK^+\pi^-$ and the measurement of its branching ratio with respect to the CF decay $\Lambda_c^+\to pK^-\pi^+$, using $980~{\rm fb}^{-1}$ of data~\cite{Yang:2015ytm}.

\begin{figure}[htb!]
\begin{center}
\includegraphics[width=0.495\textwidth, height=5.0cm]{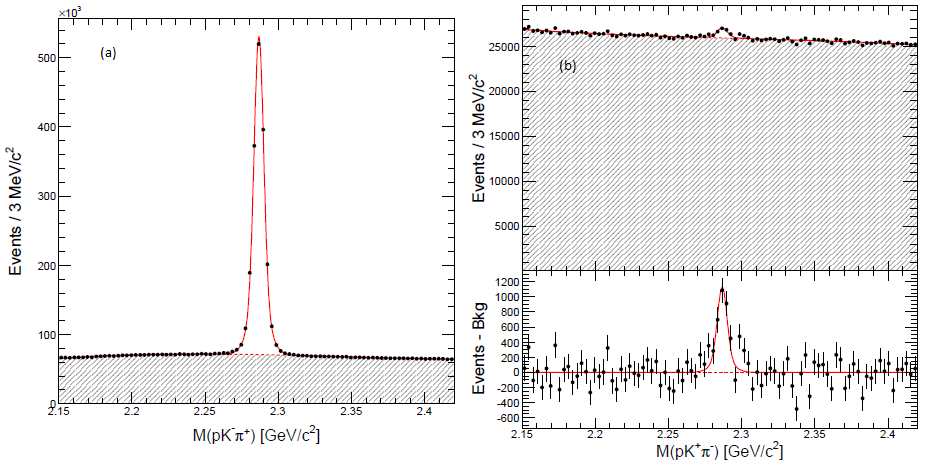}
\vskip -0.3cm
\caption{\small Distributions of (a) $M(pK^-\pi^+)$ and (b) $M(pK^+\pi^-)$ and residuals of data with respect to the fitted combinatorial background. The solid curves indicate the full fit model  and the dashed curves the combinatorial background.}
\label{fig:dcs}
\end{center}
\end{figure}

Figure~\ref{fig:dcs} shows the invariant mass distributions of (a) $pK^-\pi^+$ (CF) and (b) $pK^+\pi^-$ (DCS) combinations. DCS decay events are clearly observed in $M(pK^+\pi^-)$. In order to obtain the signal yield, a binned least-$\chi^2$ fit is performed. From the mass fit, we extract $(1.452\pm0.015)\times10^6$ $\Lambda_c^+\to pK^-\pi^+$ events and $3587\pm380$ $\Lambda_c^+\to pK^+\pi^-$ events. The latter has a peaking background from the single Cabibbo-suppressed (SCS) decay $\Lambda_c^+\to\Lambda(\to p\pi^-)K^+$, which has the same final-state topology. After subtracting the SCS contribution, we have $3379\pm380\pm78$ DCS events, where the first uncertainty is statistical and the second is the systematic due to SCS subtraction. The corresponding statistical significance is 9.4 standard deviations. We measure the branching ratio,
\begin{eqnarray*}
\frac{\mathcal{B}(\Lambda_c^+\to pK^+\pi^-)}{\mathcal{B}(\Lambda_c^+\to pK^-\pi^+)}=(2.35\pm0.27\pm0.21)\times10^{-3}, 
\end{eqnarray*}
where the uncertainties are statistical and systematic, respectively.  This measured branching ratio corresponds to $(0.82\pm0.21)\tan^4\theta_c$, where the uncertainty is the total, which is consistent  within 1.5 standard deviations with the 
na{\"i}ve expectation ($\sim\tan^4\theta_c$~\cite{Link:2005ym}). LHCb's recent measurement of $\frac{\mathcal{B}(\Lambda_c^+\to pK^+\pi^-)}{\mathcal{B}(\Lambda_c^+\to pK^-\pi^+)}=(1.65\pm0.15\pm0.05)\times10^{-3}$~\cite{Aaij:2017rin} is lower than our ratio at the 2.0$\sigma$ level.  Multiplying this ratio with the previously measured $\mathcal{B}(\Lambda_c^+\to pK^-\pi^+)=(6.84\pm0.24^{+0.21}_{-0.27})\%$ by the Belle Collaboration~\cite{Zupanc:2013iki}, we obtain the 
 the absolute branching fraction of the DCS decay, 
 \begin{eqnarray*}
 \mathcal{B}(\Lambda_c^+\to pK^+\pi^-)=(1.61\pm0.23^{+0.07}_{-0.08})\times10^{-4},
 \end{eqnarray*}
where the first uncertainty is due to the total uncertainty of the branching ratio and the second is uncertainty due to the branching fraction of the CF decay. 
After subtracting the contributions of $\Lambda^*(1520)$ and $\Delta$ isobar intermediates, which contribute only to the CF decay, the revised ratio,  $\frac{\mathcal{B}(\Lambda_c^+\to pK^+\pi^-)}{\mathcal{B}(\Lambda_c^+\to pK^-\pi^+)}=(1.10\pm0.17)\tan^4\theta_c$ is consistent with the na{\"i}ve expectation within 1.0 standard deviation.

\section{$\phi_3$ measurement with a model-independent Dalitz plot analysis of $B^{\pm}\to DK^{\pm},D\to K_S^0\pi^+\pi^-$ decay}
\begin{figure}[htb!]
\begin{center}
\includegraphics[width=0.5\textwidth]{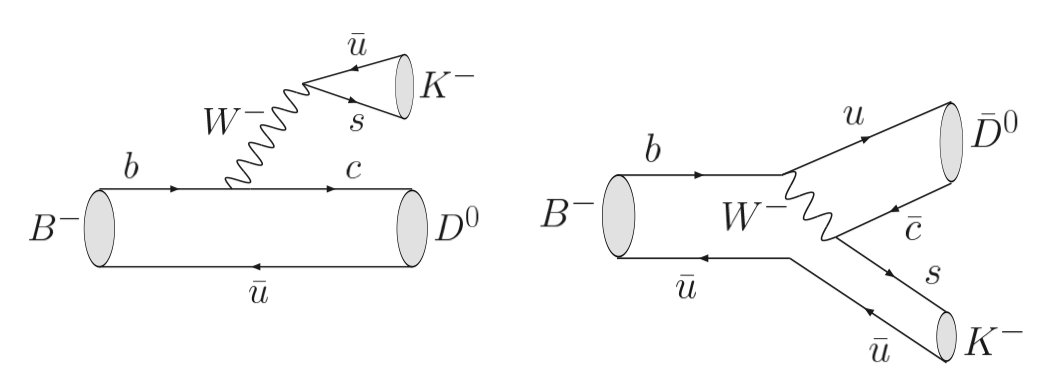}
\vskip -0.3cm
\caption{\small Feynman diagram for $B^-\to D^0K^-$ and $B^-\to\bar{D}^0K^-$ decays.}
\label{fig:phi3_fn}
\end{center}
\end{figure}

The CKM angle $\phi_3$ (also denoted as $\gamma$) is one of the least constrained parameters of the CKM Unitary Triangle. Its determination is however theoretically clean due to absence of loop contributions; $\phi_3$ can be determined using tree-level processes only, exploiting the interference between $b\to u\bar{c}s$ and $b\to c\bar{u}s$ transitions that occurs when  a process involves a neutral $D$ meson reconstructed in a final state accessible to both $D^0$ and $\bar{D}^0$  decays (see Fig.~\ref{fig:phi3_fn}). Therefore, the angle $\phi_3$ provides a SM benchmark, and its precise measurement is crucial in order to disentangle non-SM contributions to other processes, via global CKM fits. The size of the interference also depends on the ratio ($r_B$) of the magnitudes of the two tree diagrams involved and $\delta_B$, the strong phase difference between them. Those hadronic parameters will be extracted from data together with the angle $\phi_3$.

The measurement are performed in three different ways: (a) by utilizing decays of $D$ mesons to $CP$ eigenstates, such as $\pi^+\pi^-$, $K^+K^-$ ($CP$ even) or $K_S^0\pi^0$, $\phi K_S^0$ ($CP$ odd), proposed by M. Gronau, D. London, and D. Wyler (and called the GLW method~\cite{Gronau:1990ra, Gronau:1991dp})(b) by making use of DCS decays of $D$ mesons, $e.g.$, $D^0\to K^+\pi^-$, proposed by D. Atwood, I. Dunietz, and A. Soni (and called the ADS method~\cite{Atwood:2000ck}) and (c) by exploiting the interference pattern in the Dalitz plot of the $D$ decays such as $D^0\to K_S^0\pi^+\pi^-$, proposed by A. Giri, Y. Grossman, A. Soffer, and J. Zupanc (and called the GGSZ method~\cite{Giri:2003ty}). 

Using a model-dependent Dalitz plot method, Belle's earlier measurement~\cite{Poluektov:2010wz}  based on a data sample of $605~{\rm fb^{-1}}$ integrated luminosity yielded $\phi_3=(78.4^{+10.8}_{-11.6}\pm3.6\pm8.9)^{\circ}$ and $r_B=0.160^{+0.040}_{-0.038}\pm0.011^{+0.050}_{-0.010}$, where the uncertainties are statistical, systematic and  Dalitz model dependence, respectively. Although with more data one can squeeze on the statistical part, the result will still remain limited by the model uncertainty.

In a bid to circumvent this problem, Belle has carried out a model-independent analysis~\cite{Aihara:2012aw}, using GGSZ method~\cite{Giri:2003ty},  that is further extended in a latter work~\cite{Bondar:2008hh}. The analysis is based on the $711~{\rm fb^{-1}}$ of data, collected at the $\Upsilon(4S)$ resonance. In contrast to the conventional Dalitz method, where the $D^0\to K_S^0\pi^+\pi^-$ amplitudes are parameterized as a coherent sum of several quasi two-body amplitudes as well as a nonresonant term, the model-independent approach invokes study of a binned Dalitz plot. In this approach, the expected number of events in the $i^{th}$ bin of the Dalitz plan for the $D$ mesons from $B^{\pm}\to DK^{\pm}$ is given by
\begin{equation}
N^{\pm}_i=h_B\big[K_{\pm i}+ r^2_BK_{\mp i}+2\sqrt{K_iK_{-i}}(x_{\pm}c_i\pm y_{\pm}s_i)\big],
\end{equation}
where $h_B$ is the overall normalization and $K_i$ is the number of events in the $i^{th}$ Dalitz bin of the flavor-tagged (whether $D^0$ or $\bar{D}^0$) $D^0\to K_S^0\pi^+\pi^-$ decays, accessible via the charge of the slow pion in $D^{*\pm}\to D\pi^{\pm}$. The terms $c_i$  and $s_i$  contain information about the strong-phase difference between the symmetric Dalitz points [$m^2(K_S^0\pi^+),~m^2(K_S^0\pi^-)$] and [$m^2(K_S^0\pi^-),~m^2(K_S^0\pi^+)$]; they are the external inputs obtained from quantum correlated $D^0\bar{D^0}$ decays at the $\psi(3770)$ resonance in CLEO~\cite{Briere:2009aa, Libby:2010nu}. Finally $x_{\pm} = r_B \cos(\delta_B \pm \phi_3)$ and $y_{\pm} = r_B \sin(\delta_B \pm \phi_3)$, where $\delta_B$ is the strong-phase difference between $B^{\pm}\to \bar{D}^0K^{\pm}$ and $B^{\pm}\to D^0K^{\pm}$.

We perform a combined likelihood fit  to four signal selection variables in all Dalitz bins (16 bins in our case) for 
the  $B^{\pm}\to DK^{\pm}$ signal and Cabibbo-favored $B^{\pm}\to D\pi^{\pm}$ control samples; the free parameters of the fit are $x_{\pm}$, $y_{\pm}$, overall normalization (see Eq. 1) and background fraction. Table~\ref{tab:xy} summarizes
the results obtained for $B^{\pm}\to DK^{\pm}$ decays. From these results, we obtain $\phi_3 = (77.3^{+15.1}_{-14.9}\pm 4.1 \pm 4.3)^{\circ}$ and $r_B = 0.145\pm0.030\pm0.010\pm0.011$, where the first error is statistical, the second is systematic, and the last error is due to limited precision on $c_i$ and $s_i$. Although $\phi_3$ has a mirror solution at $\phi_3 + 180^{\circ}$, we retain the value consistent with $0^{\circ} < \phi_3 <180^{\circ}$. 
\begin{table}
\centering
\caption{\small Results of  the $x, y$ parameters and their statistical correlation for $B^{\pm}\to DK^{\pm}$ decays. The quoted uncertainties are statistical, systematic, and precision on $c_i$, $s_i$, respectively.}
\begin{tabular}{c|c}
\hline\hline
   Parameter & \\
\hline
$x_{+}$ &  $+0.095\pm0.045\pm0.014\pm0.010$\\
$y_{+}$  &  $+0.135^{+0.053}_{-0.057}\pm0.015\pm0.023$\\
corr($x_{+}$, $y_{+}$) & -0.315\\
$x_{-}$ & $-0.110\pm0.043\pm0.014\pm0.007$\\
$y_{-}$ & $-0.050^{+0.052}_{-0.055}\pm0.011\pm0.017$\\
corr($x_{-}$, $y_{-}$) & +0.059\\
\hline    \hline    
\end{tabular}
\label{tab:xy}
\end{table}
We report evidence for direct $CP$ violation, the fact that $\phi_3$ is nonzero, at the 2.7 standard deviations  level. Compared to results of the model-dependent Dalitz method, this measurement has somewhat poorer statistical precision despite a larger data sample used. There are two factors responsible for lower statistical sensitivity: 1) the statistical error for the same statistics is inversely proportional to the $r_B$ value, and the central value of $r_B$ in this analysis is smaller, and 2) the binned approach is expected to have the statistical precision that is, on average, 10--20\% poorer than the unbinned one. On the positive side, however, the large model uncertainty for the model-dependent study ($8.9^{\circ}$) is now replaced by a purely statistical uncertainty due to limited size of the $\psi(3770)$ data sample available at CLEO ($4.3^{\circ}$). With the use of BES-III data, this error will decrease to $1^{\circ}$ or less. 
    
The model-independent approach therefore offers an ideal avenue for Belle II and LHCb  in their pursuits of $\phi_3$. We expect that the statistical error of the $\phi_3$ measurement using the statistics of a $50~{\rm ab^{-1}}$ data sample that will be available at Belle II will reach $1-2^{\circ}$. We also expect that the experimental systematic error can be kept at the level below $1^{\circ}$, since most of its sources are limited by the statistics of the control channels.
\\
\\
\textbf{Acknowledgement}:
The author thanks the organizers of PKI2018 for excellent hospitality and for assembling a nice scientific program. This work is supported by the U.S. Department of Energy.

\end{document}